\documentclass[sigconf]{acmart}
\usepackage{pgfplots}

\AtBeginDocument{%
  \providecommand\BibTeX{{%
    \normalfont B\kern-0.5em{\scshape i\kern-0.25em b}\kern-0.8em\TeX}}}

\renewcommand\footnotetextcopyrightpermission[1]{} 
\begin{document}
\usetikzlibrary{patterns}


\title{Sequential Deep Learning for Credit Risk Monitoring with Tabular Financial Data}


\author{Jillian M. Clements}
\affiliation{%
    \institution{American Express, AI Research}
    \city{New York}
    \state{NY}
    }
\email{jillian.clements@aexp.com}

\author{Di Xu}
\affiliation{%
    \institution{American Express, AI Research}
    \city{New York}
    \state{NY}
    }
\email{di.w.xu@aexp.com}

\author{Nooshin Yousefi}
\affiliation{%
    \institution{Rutgers University}
    \city{New Brunswik}
    \state{NJ}
    }
\email{no.yousefi@rutgers.edu}

\author{Dmitry Efimov}
\affiliation{%
    \institution{American Express, AI Research}
    \city{New York}
    \state{NY}
    }
\email{dmitry.efimov@aexp.com}

\renewcommand{\shortauthors}{Clements, Xu, Yousefi, and Efimov}

\begin{abstract}
Machine learning plays an essential role in preventing financial losses in the banking industry. Perhaps the most pertinent prediction task that can result in billions of dollars in losses each year is the assessment of credit risk (i.e., the risk of default on debt). Today, much of the gains from machine learning to predict credit risk are driven by gradient boosted decision tree models. However, these gains begin to plateau without the addition of expensive new data sources or highly engineered features. In this paper, we present our attempts to create a novel approach to assessing credit risk using deep learning that does not rely on new model inputs. We propose a new credit card transaction sampling technique to use with deep recurrent and causal convolution-based neural networks that exploits long historical sequences of financial data without costly resource requirements. We show that our sequential deep learning approach using a temporal convolutional network outperformed the benchmark non-sequential tree-based model, achieving significant financial savings and earlier detection of credit risk. We also demonstrate the potential for our approach to be used in a production environment, where our sampling technique allows for sequences to be stored efficiently in memory and used for fast online learning and inference.
\end{abstract}






\maketitle
\pagestyle{plain}

\section{Introduction}
When bank borrowers get into trouble, the early detection of small changes in their financial situation can substantially reduce a bank's losses when precautionary steps are taken (e.g., lowering the borrower's credit limit). In fact, the ongoing monitoring of credit risk is an integral part of most loss mitigation strategies within the banking industry \cite{thomas2002}. 

Current credit scoring models to assess risk typically rely on gradient boosted decision trees (GBDTs), which provide several benefits over more traditional techniques such as logistic regression \cite{Friedman2001}. For instance, interactions between variables are automatically created during the boosting process and provide multiple combinations to improve prediction accuracy. Additionally, preprocessing steps such as missing value imputation and data transformation are generally not required. However, these models also have certain limitations. First, performance improvements are usually derived from feature engineering to create new variables based on extensive domain knowledge and expertise. Second, tree-based models do not take full advantage of the available historical data. Finally, they cannot be used in an online learning setting (where data becomes available in a sequential order) due to their limited capacity for updates, such as altering splits, without seeing the whole dataset from the start (e.g., \cite{chen2016}).

To overcome the limitations of a traditional tree-based approach, we explored several deep learning techniques for assessing credit risk and created a unique method for generating sequences of credit card transactions that looks back one year into borrowers' financial history. Applying the same input features as the benchmark GBDT model, we show that our final sequential deep learning approach using a temporal convolution network (TCN) provides distinct advantages over the tree-based technique. The main improvement is that performance increases substantially, resulting in significant financial savings and earlier detection of credit risk. We also demonstrate the potential for our approach to be used in an online learning setting for credit risk monitoring without making significant changes to the training process.

Previous research on the application of deep learning to tabular data from credit card transactions has focused on fraud detection \cite{Roy2018, Efimov2019}, loan applications \cite{Wang2018, Babaev2019}, and credit risk monitoring \cite{Nanni2009,Addo2018}. Our work differs from previous deep learning work on credit risk monitoring by extending beyond simple multilayer perceptron networks and into sequential techniques, such as recurrent neural networks (RNNs) and TCNs, that explicitly utilize the available historical data. Additionally, we introduce a novel sampling method for generating sequences of transactions that addresses the drastic differences in sequence length between card members over a full year. The benefits of using sampled transactions from a long time window, as opposed to the total number of transactions, are that it reduces noise, memory requirements, and training/inference time so that the model is able to learn generalizable early warning signs of risky financial behavior while still being able to run in near real time during online credit monitoring.

The main contributions presented in this paper are as follows. We first show the potential of sequential deep learning to improve our current credit scoring system through offline analysis, which revealed significant reductions in financial losses (tens of millions of US dollars annually) and increased early risk detection rates. We also developed a sampling method for generating sequences of transactions over the course of a full year, which allows the model to learn long-term behavioral trends and removes the need to load and process the hundreds of billions of credit card transactions that occur in a given year across card members. Finally, we demonstrate that online learning can be applied to our approach and achieves higher performance over randomly re-initializing the model weights.

\section{Methods}

\subsection{Transactional Data}
Credit risk scores are predicted using tabular data from credit card transactions, where each transaction is associated with a set of numerical and categorical variables.  We define credit risk prediction as a \textit{binary classification} task, where the goal is to detect default (non-payment) on credit card debt within 1.5 years following the timestamp of each transaction.

\subsubsection{Features}
For our modeling experiments, we used transactional data for card members with consumer credit cards. Data sources included both internal and external sources, such as consumer credit reporting agencies, from which 127 raw and engineered features were created. Our features can be divided into two subcategories: transaction-related (credit exposure, ATM indicator, etc.) and account-related (days past due, account tenure, etc.). To ensure fairness, personal demographic information was not included in any of the features. However, due to privacy concerns and strict adherence to data protection laws, we are unable to describe our features in further detail. 

\subsubsection{Training and Validation}
The training dataset consisted of a sample of 15 million card members, with transactions spanning the course of twelve months from March 2016 to February 2017. We shifted the one year window forward by one month twice to create 45 million transactional sequences for training. This simply means that each card member appeared three times in the training dataset with slightly different (time-shifted) sequence data.

The validation dataset consisted of a sample of 6 million non-overlapping (out-of-sample) card members, with transactions spanning twelve months from May 2017 to April 2018 in order to also create an out-of-time validation set. This simulates our production environment, where models are trained on past data. 

Within these sample datasets, roughly 2\% of card members defaulted on their balance. We addressed this class imbalance in the context of deep learning by creating balanced mini-batches for training models with mini-batch gradient descent, where each mini-batch had the same percentage of defaulting card members as the whole dataset.

\subsubsection{Preprocessing}
Neural networks are sensitive to feature distributions and performance can drop significantly for non-normal distributions, such as skewed distributions or those with outliers. Our tabular datasets contain many features with similar behavior and thus, standard feature preprocessing approaches would not lead to superior results. Therefore, our preprocessing steps were as follows.
\begin{itemize}
\item Missing values for each feature were imputed based on the target label (default/non-default) rates within 10 bins defined by the feature's percentiles. 
\item For dealing with numerical outliers, we developed a novel capping procedure to extract the most significant part of the feature distribution by leveraging splits obtained from training a decision tree model (in our case, the GBDT framework was used). For each feature and all trees produced by the model, we sorted all splits in ascending order:
$$
(s_1, s_2, \ldots, s_k), \, s_i > s_j, \mbox{ for all } i>j
$$
and applied the following capping rules:
$$
\hat{x} = \left\{
\begin{array}{ll}
2 \cdot s_1 - s_2, & \mbox{if } x < 2 \cdot s_1 - s_2, \\
2 \cdot s_k - s_{k-1}, & \mbox{if } x > 2 \cdot s_k - s_{k-1}, \\
x, & \mbox{otherwise},
\end{array}
\right.
$$
where $x$ is the original feature and $\hat{x}$ is the transformed feature (see Figure \ref{fig:xgbcap} for an illustrative example).

\item After imputation and outlier capping, the Box-Cox transformation was applied to reduce feature skewness \cite{Box1964}. 
\item Categorical data were transformed to numerical features using a procedure known as Laplace smoothing \cite{Manning2008}, which contains two main steps:
\begin{enumerate}
    \item Calculate the average of the target variable within each category.
    \item Modify the average from step 1 to address categories with a small number of observations:
    $$
    \hat{x} = \dfrac{k \cdot \bar{y} + \sum\limits_{i \in G} y_i}{k + \left|G\right|},
    $$
    where $G$ is a set of indices for the given category, $|G|$ is the size of the category, $k$ is a metaparameter defined empirically ($k = 30$ in our case), and $\bar{y}$ is the average of the target variable for all training observations.
\end{enumerate}
\begin{figure}
  \centering
  \includegraphics[width=0.45\textwidth]{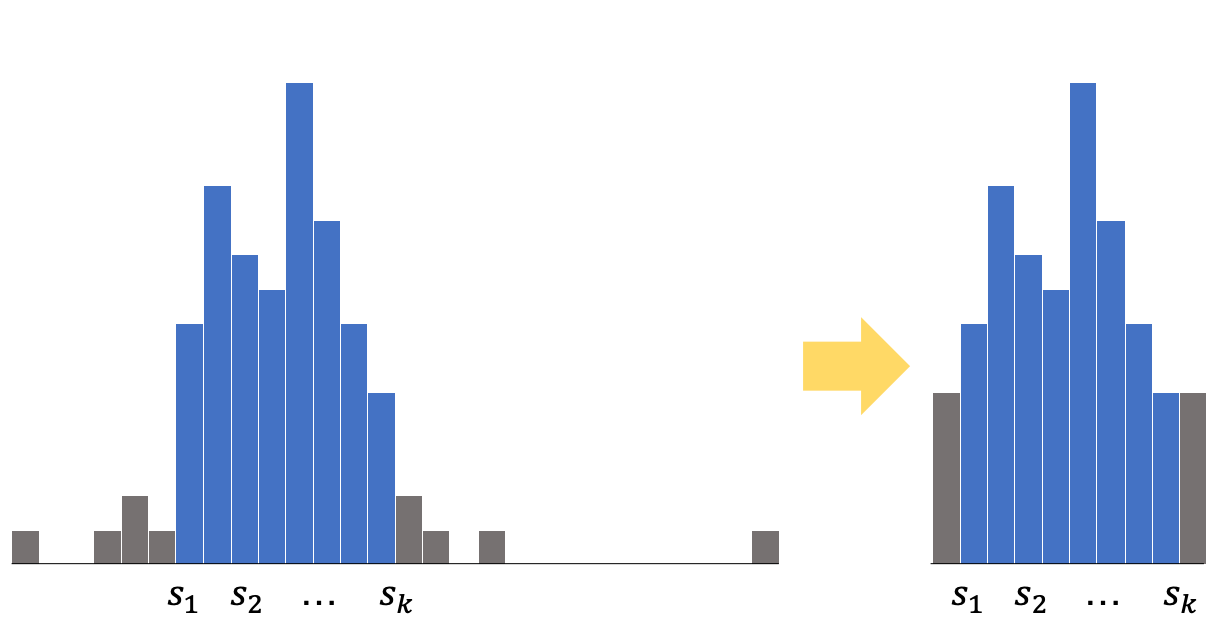}
  \caption{Capping with GBDT splits.}
  \label{fig:xgbcap}
\end{figure}
\item Finally, standard scaling was applied to all features. \end{itemize}
 
\subsection{Sampling Transactions for Sequence Generation}
Generating sequences of credit card transactions over a fixed period of time presents a unique challenge for risk assessment. Within a given time frame, card members can have drastically different amounts of transactions. For example, in a single year, one card member might have only a few transactions while another has thousands. Therefore, rather than modeling full sequences of transactions over the course of year, we created a sampling scheme that selects one random historical transaction per card member per month. 

\begin{figure}
  \centering
  \includegraphics[width=0.475\textwidth]{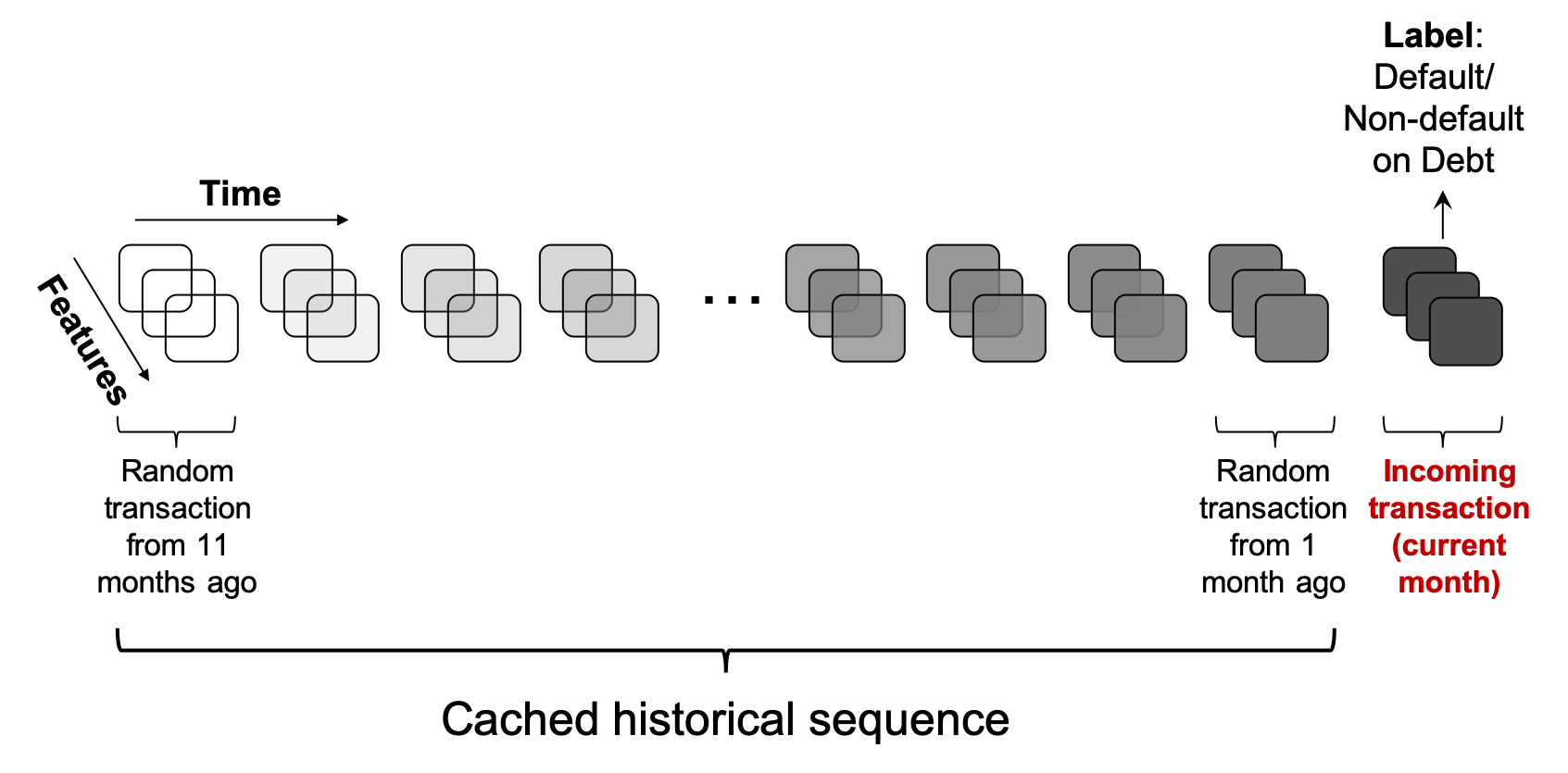}
  \caption{Sequence of sampled transactions.}
  \label{fig:seq}
\end{figure}

The benefit of this sampling technique is twofold. First, it reduces noise in order to expose more general trends in risky card member behavior over a long period of time. Second, it allows us to create and process sequences in near real time because the previous 11 months of transactions can be stored efficiently in memory and loaded quickly. We simply append each incoming transaction in the current month to the end of the sequence before submitting it to a sequential model (see Figure \ref{fig:seq} for an illustration). This is important for the implementation of sequential models in our production environment, where a near real-time processing speed is needed to calculate a credit risk score for each incoming transaction before the transaction is approved (usually <10 milliseconds).

For inactive card members with zero transactions in a given month, data were collected at the timestamp of their monthly billing statement date. The reason we use random historical transactions instead of the billing statement date for active card members is because we want to predict credit risk for each incoming transaction, which does not occur at a fixed time interval. Therefore, we introduce noise to the time interval in order to reduce the model's dependence on it.

For low tenure card members with less than 12 months of transactions available, we zero-padded the sequences and applied a binary mask to the loss calculations during training. This technique excludes the padded timestamps from being used to update the model weights.

\subsection{Model Evolution}
We compared four different types of deep learning classification models, from simple neural networks to more complex ones, and compared their performance to a benchmark GBDT model for predicting credit risk. Since the GBDT algorithm was not designed to accommodate sequential data, we only used the last (i.e., most recent) month's transaction to train the model. Similarly, two of the deep learning models (multilayer perceptron and TabNet) are also non-sequential and only used the most recent month's transaction.


\subsubsection{Multilayer Perceptron}
We started our deep learning experiments with a straightforward implementation of the standard vanilla multilayer perceptron (MLP) neural network with dropout regularization. The MLP connects multiple layers of nodes/neurons in a forward-directed graph. Dropout regularization stochastically "drops" some of these neurons during training in order to simulate an ensemble of different MLPs \cite{Hinton2012}. To train the connection weights of the model, a generalization of the least mean squares algorithm known as backpropagation was used to calculate the gradient of the loss function with respect to each weight by the chain rule, iterating backward from the last layer.

\subsubsection{TabNet} 
\begin{figure}
  \centering
  \includegraphics[width=0.4\textwidth]{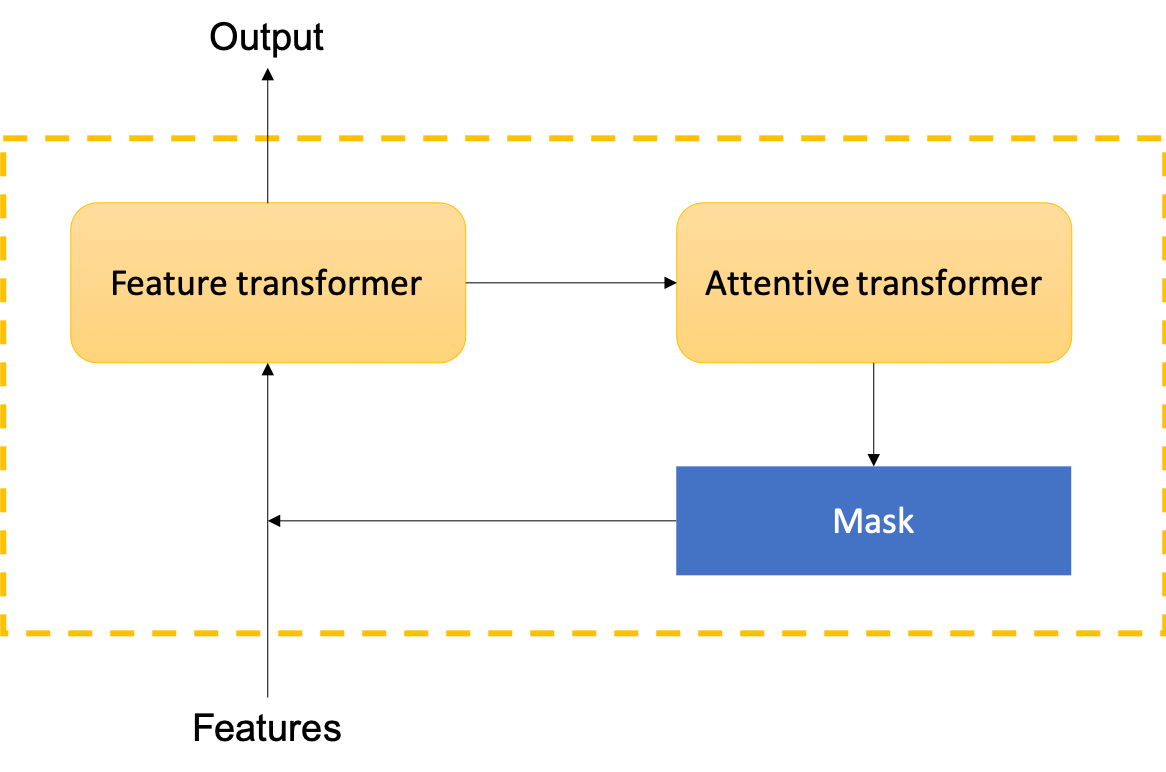}
  \caption{TabNet's decision step architecture.}
  \label{fig:tabnet}
\end{figure}

TabNet is a more recently developed neural network designed specifically for tabular, non-sequential data \cite{Arik2019}. It utilizes an iterative attention mechanism for feature selection, where the final prediction score is an aggregate of all of the processed information from the previous iterations. A single iteration of the TabNet architecture contains two parts: a feature transformer and an attentive transformer. The output of the feature transformer is combined with the outputs from the previous iterations in order to obtain a final output decision. The attentive transformer creates a sparse feature mask vector, which is applied back to the initial feature set to generate a smaller subset of features that is fed into the next decision step (see Figure \ref{fig:tabnet}). The benefit of this technique is that it allows the network to fully utilize the most relevant features in a tabular dataset at each decision step, which enables more efficient learning.

\subsubsection{Recurrent Neural Network} The aforementioned deep learning models suffer from a similar limitation as the baseline GBDT model: they don't explicitly utilize card members' historical data. RNNs address this issue by iteratively processing and saving information from previous transactions in its hidden nodes, which is then added to the current transaction in order to produce a more informed credit risk prediction. 

\begin{figure}
  \centering
  \includegraphics[width=0.45\textwidth]{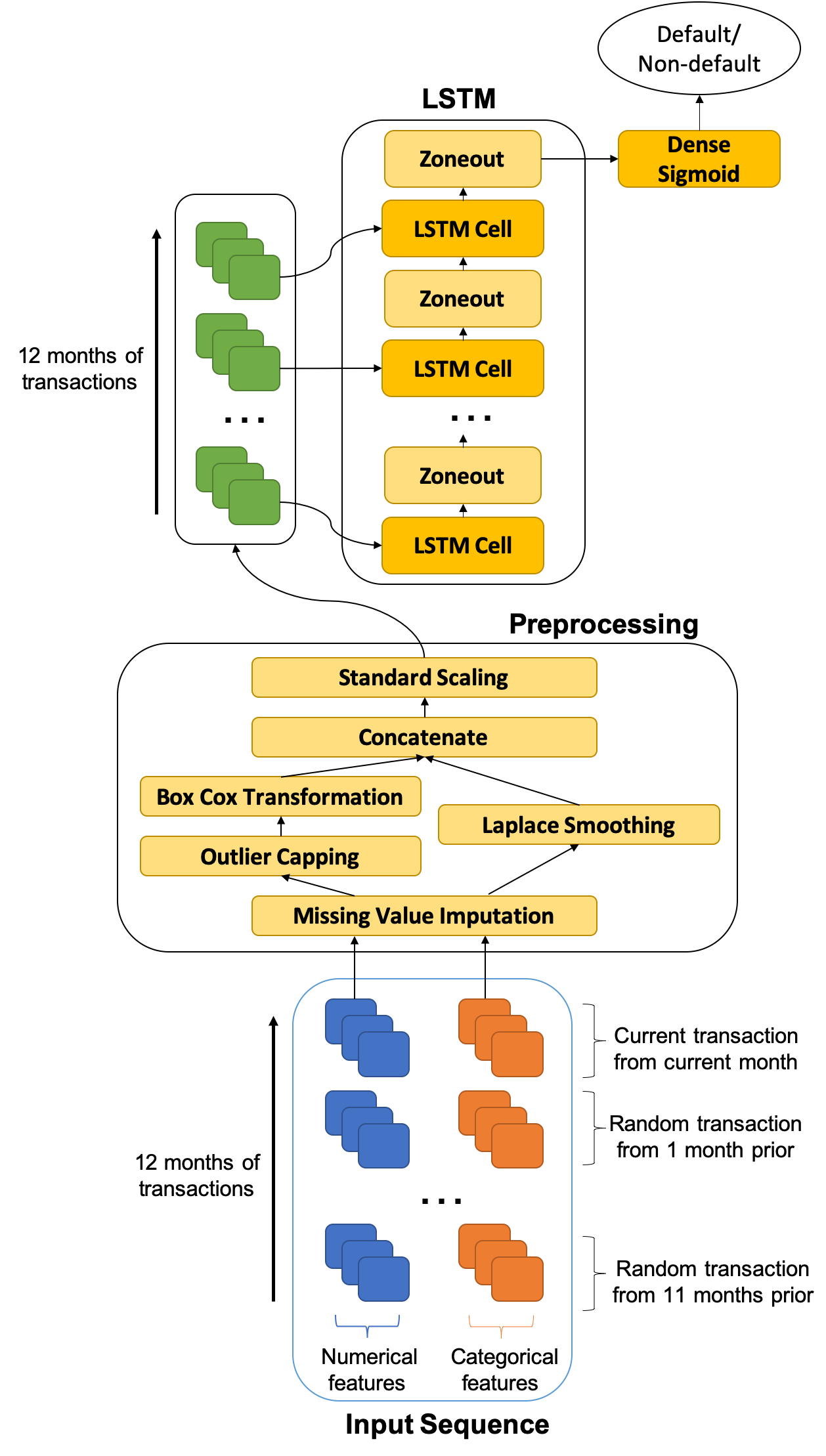}
  \caption{Single-layer LSTM architecture for predicting credit risk (i.e., default on credit card debt) from 12 months of transactional data.}
  \label{fig:rnn}
\end{figure}

For this paper, we used a long short-term memory (LSTM) RNN \cite{Hochreiter1997}, which is a popular variant of the vanilla RNN that has been shown to more effectively learn long-term dependencies in sequential data. We also utilized a newer method for RNN regularization known as \textit{zoneout regularization} \cite{Krueger2019}. Like dropout regularization, zoneout uses random noise to approximate training an ensemble of different RNNs. However, instead of dropping random recurrent connection weights, zoneout stochastically forces some of the hidden nodes to maintain their previous values from the previous transaction. The benefit of using zoneout over recurrent dropout is that it allows the RNN to remember more information from past transactions. An example of a single-layer LSTM model with zoneout regularization for predicting credit risk scores using our preprocessing approach is shown in Figure \ref{fig:rnn}.

\begin{figure}
  \centering
  \includegraphics[width=.475\textwidth]{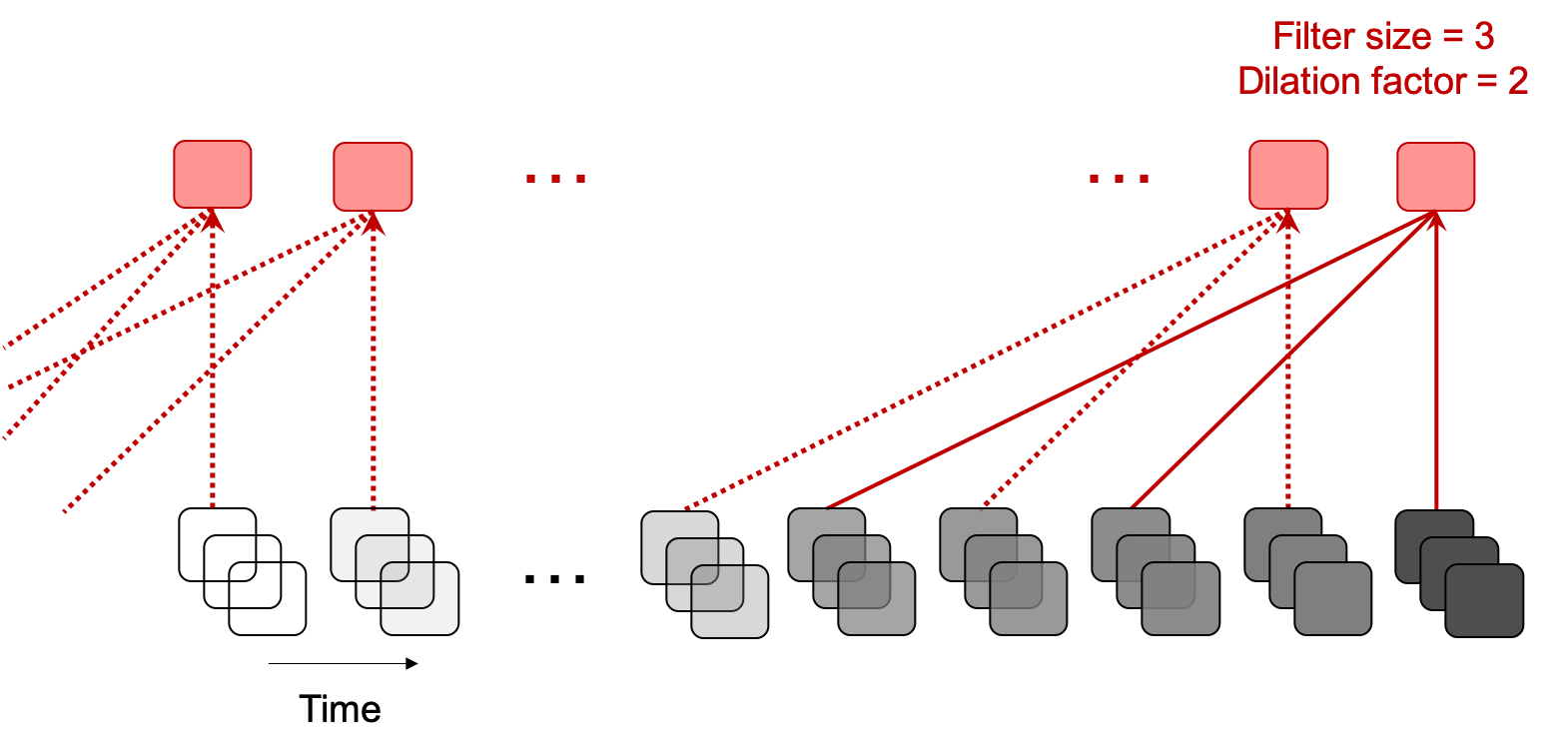}
  \caption{Example of 1D causal convolution.}
  \label{fig:causalconv}
  \vspace*{\floatsep}
  \includegraphics[width=.3\textwidth]{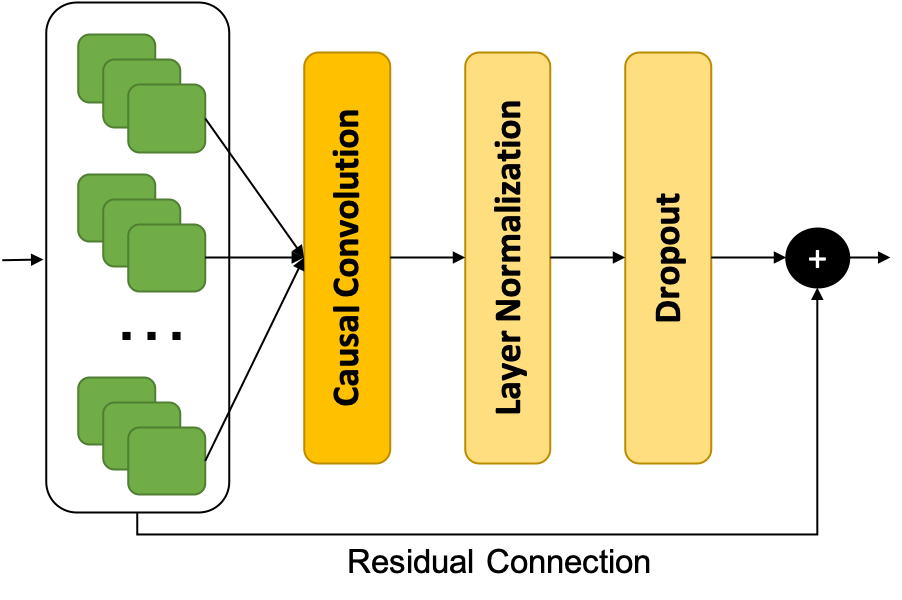}
  \caption{TCN block.}
  \label{fig:tcnblock}
  \vspace*{\floatsep}
  \includegraphics[width=.475\textwidth]{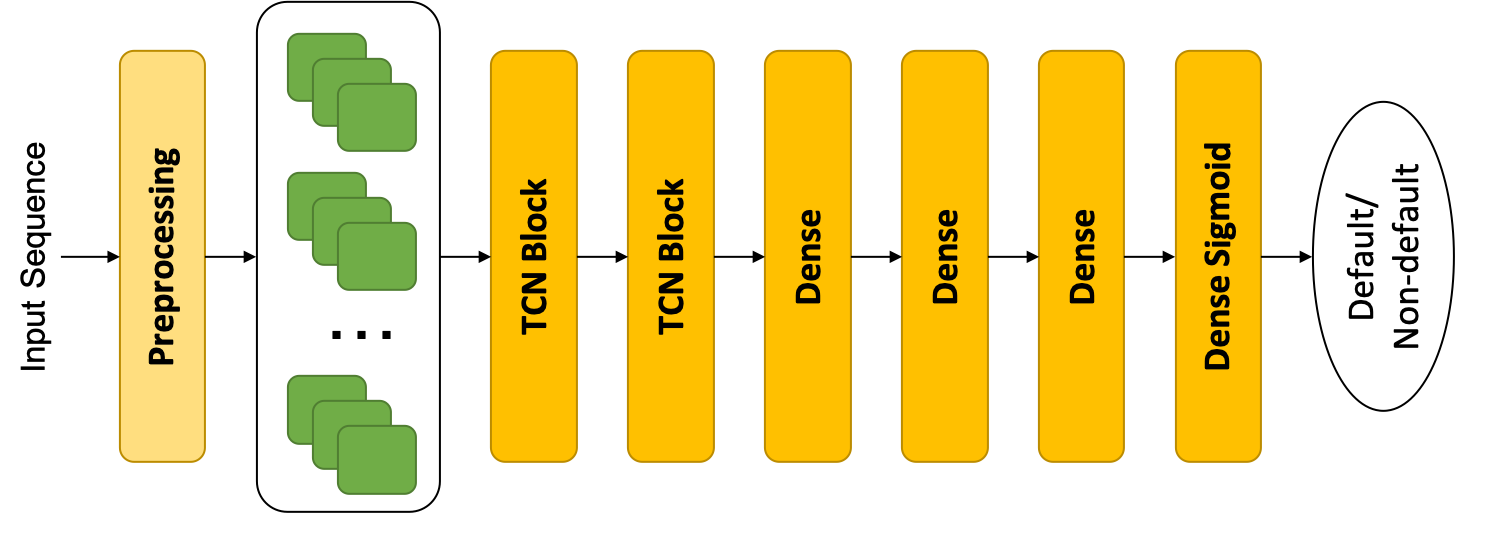}
  \caption{Final TCN architecture.}
  \label{fig:tcn}
\end{figure}

\subsubsection{Temporal Convolutional Network} An important limitation of the RNN framework is that training the model can be time-consuming because sequences of data are processed iteratively. Additionally, the ability of RNNs to capture long-term dependencies in sequences remains a fundamental challenge \cite{pascanu2013}. Therefore, more recent techniques have focused on more efficient convolution-based approaches for sequential data \cite{oord2016, kalchbrenner2016, dauphin2017, gehring2017, Bai2018}. Since convolutions can be done in parallel, sequences are processed as a whole instead of iteratively as in RNNs. Convolutional neurons can also utilize their receptive field to retain even more information from the distant past. \textit{Causal} convolutions prevent leakage from the future by only convolving the output at time \textit{t} with elements from time \textit{t} and earlier (see Figure \ref{fig:causalconv}) \cite{waibel1989}.

Perhaps the most popular recent causal convolution-based approach is the temporal convolutional network (TCN) \cite{Bai2018}. In addition to convolution and dropout, TCNs utilize dilation to enable larger receptive field sizes that look back at longer history lengths by introducing a fixed step size between every two adjacent convolutional filter taps \cite{Yu2016}. TCNs also employ residual connections so that each convolutional layer learns an identity mapping rather than the entire transformation (see Figure \ref{fig:tcnblock}), which helps stabilize deeper and wider models \cite{He2016}.  Within each residual TCN block, layer normalization is applied to the output of the convolutional layer and dropout is added for regularization. An illustration of our final deep TCN architecture for the credit risk prediction task is shown in Figure \ref{fig:tcn}.

\subsection{Model Optimization and Tuning}
\subsubsection{Architecture Search} The final number of layers, neurons per layer, and dropout/zoneout rates used in each deep learning model were determined using an iterative Bayesian optimization approach with SigOpt \cite{Dewancker2015}. However, several of the top models performed similarly. We found that the best performance occurred for the 5-layer MLP, 3-layer LSTM (with 2 LSTM layers and one dense output layer), and 6-layer TCN (with 2 TCN blocks and 4 dense layers). Dropout and zoneout rates were kept fairly low in each model (averaging \textasciitilde0.2).

\subsubsection{Optimization}
Each deep learning model was trained using the Adam optimization algorithm \cite{Kingma2014} and early stopping \cite{prechelt1998}. We used the default beta parameters for the Adam algorithm (betas = 0.9, 0.999) that control the decay rate for the first and second moment estimates of the gradient, although other values were explored and showed no obvious improvement. The classic binary cross entropy function for classification tasks was used as the loss function to be optimized.

\subsubsection{Batch Size and Learning Rate} We also experimented with several batch sizes, learning rates, and learning rate schedules for the different architectures. We found that a universal batch size of 512 credit card members and a decay rate of 0.8, starting from an initial learning rate of 1e-4, was generally the most effective for all models.

\subsubsection{Performance Metrics} Performance was measured in terms of the Gini coefficient and recall. The Gini coefficient is a common measurement of the discriminatory power of rating systems such as credit scoring models and is directly proportional to the area under the receiver operator characteristic curve (Gini = (2*AUROC) - 1). Recall, or sensitivity, is a measurement for the correct predictions of credit card default within a fixed fraction of the model's top prediction scores. This fraction was determined by existing business principles and fixed across models.

\section{Results}
The performance results for each individual deep learning model and its ensemble with the benchmark GBDT are shown in Table \ref{tab:performance}. Performance results for the high debt exposure subpopulation (card members with a balance exceeding \$15,000) are also included in this table to demonstrate the potential for substantial financial savings using our proposed approach.

The sequential models, LSTM and TCN, both outperformed the benchmark GBDT in isolation. The non-sequential neural networks, MLP and TabNet, performed worse than the benchmark. Additionally, a much larger boost in prediction performance was observed when the LSTM and TCN were used in an ensemble with GBDT, suggesting that historical information provides orthogonal information that is predictive of risky financial behavior. While these performance improvements may seem modest, it is important to keep in mind the large volume of card members that exists in the dataset, implying that small improvements lead to significant savings (in our case, an annual savings of tens of millions of US dollars).

\begin{table}
  \caption{Individual and ensemble model performance for the overall population and the high debt population (>\$15k).}
  \label{tab:performance}
  \begin{tabular}{c|cc|cc}
    \toprule
     &  \multicolumn{2}{c|}{Overall} &  \multicolumn{2}{c}{High Debt}\\
    Model & Gini (\%) & Recall (\%) & Gini (\%) & Recall (\%) \\
    \midrule
    GBDT & 92.19& 63.85 & 84.12 & 60.68\\
    MLP & 92.06 & 63.42 & 83.98 & 60.82\\
    TabNet & 92.03 & 63.5 & 83.92 & 61.02\\
    LSTM & 92.28 & 64.04 & 84.27 & 61.00\\
    \textbf{TCN} & \textbf{92.33}& \textbf{64.13} & \textbf{84.47} & \textbf{61.33} \\
    \midrule
    GBDT + MLP & 92.26& 64.10 & 84.44 & 61.64 \\
    GBDT + TabNet & 92.26 & 64.11 & 84.46 & 61.57 \\
    \textbf{GBDT + LSTM} & \textbf{92.49}& 64.75 & \textbf{84.83} & \textbf{62.13} \\
    \textbf{GBDT + TCN} & 92.48 & \textbf{64.81} & \textbf{84.83} & 62.01 \\
    \bottomrule 
  \end{tabular}
\end{table}

 
\begin{figure*}[h]
\begin{minipage}[t]{.45\textwidth}
  \centering
  \includegraphics[width=1\textwidth]{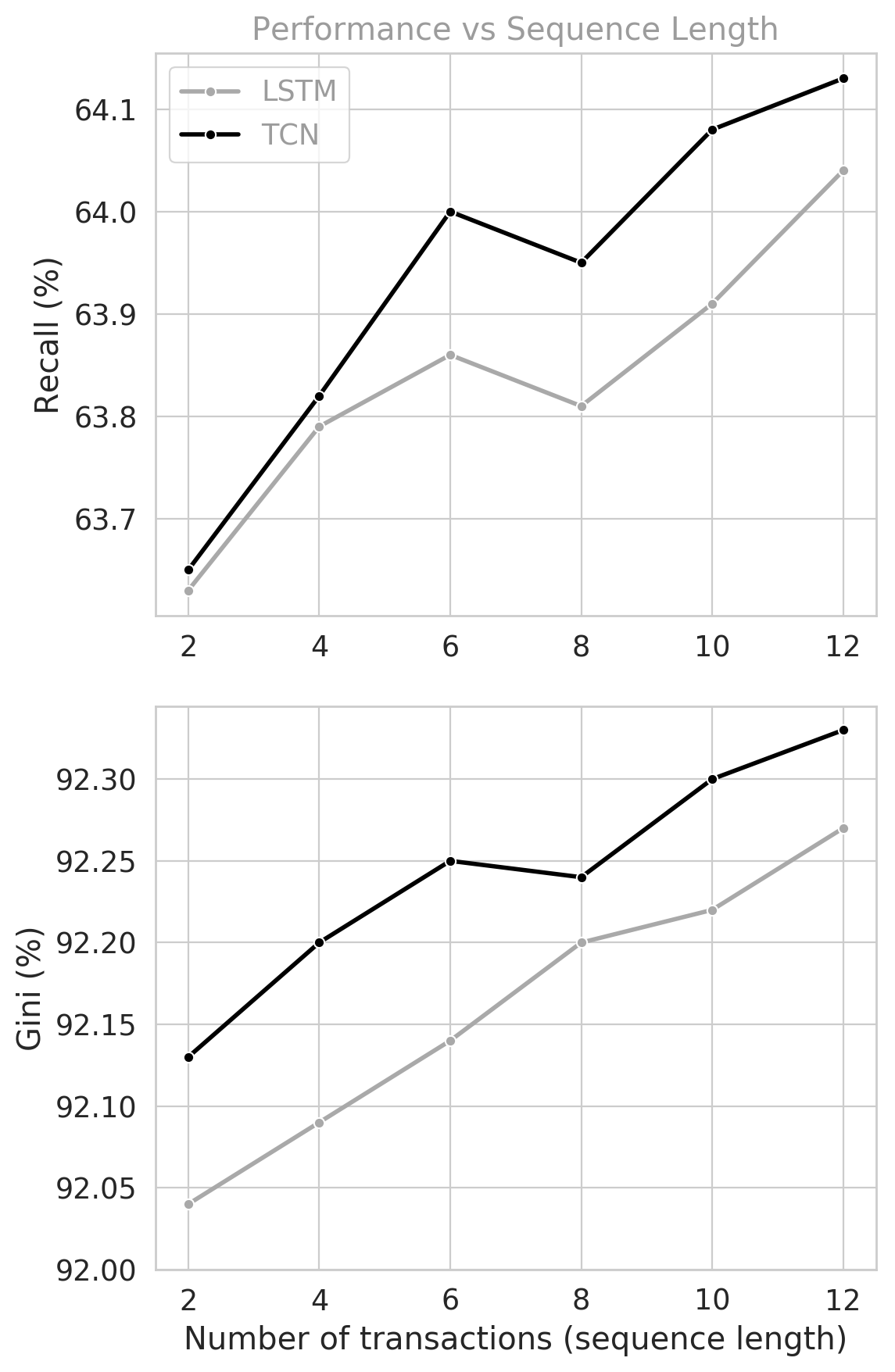}
  \caption{Performance increased as the number of monthly transactions in the sequence increased.}
  \label{fig:seqlen}
\end{minipage}\hfill
\begin{minipage}[t]{.45\textwidth}
  \centering
  \includegraphics[width=1\textwidth]{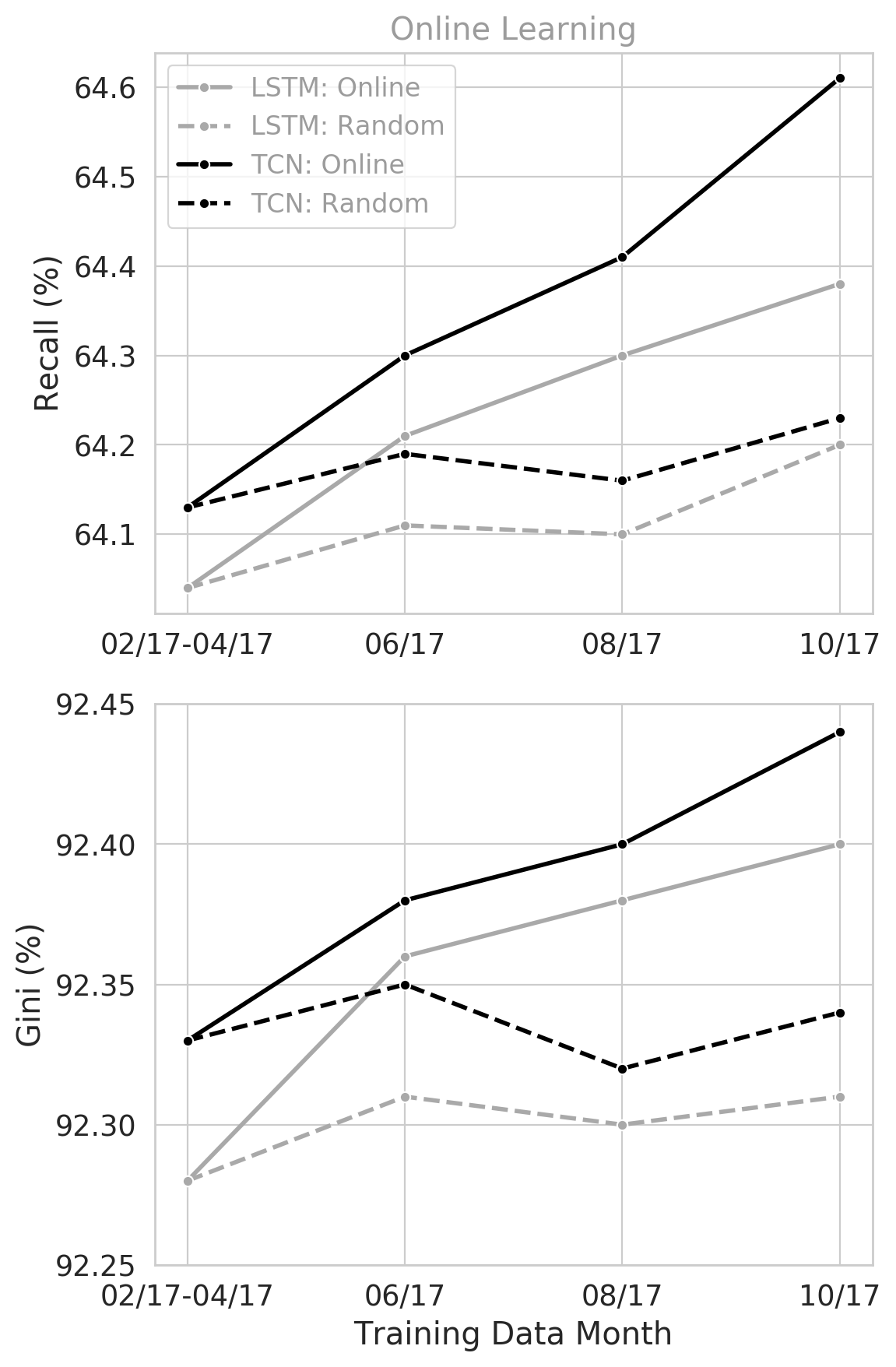}
  \caption{Online learning (i.e., progressively tuning the weights using incoming data) produced superior performance results when compared to re-initializing the weights with small random values before training.}
  \label{fig:online}
\end{minipage}\hfill
\end{figure*}

\subsection{Early Risk Detection}
In addition to capturing more risky behavior types, another important task for the model was to catch risky behavior as early as possible before the credit default occurred. This is essential when precautionary measures need to be implemented months in advance in order to reduce future financial losses. 

\begin{table}
  \caption{Individual model performance, split by default date.}
  \label{tab:bins}
  \begin{tabular}{c|c|c|c}
    \toprule
    & \multicolumn{3}{c}{Gini/Recall (\%) by Default Date} \\
     & April - & November, 2018 - & May - \\
    Model & October, 2018 & April, 2019 & November, 2019\\
    \midrule
    GBDT & \textbf{95.56 / 81.60} & 90.38 / 59.64 & 86.24 / 45.95 \\
    MLP & 95.48 / 80.90 & 90.24 / 59.26 & 86.08 / 45.81 \\
    TabNet & 95.43 / 81.08 & 90.21 / 59.26 & 86.07 / 45.90 \\
    LSTM & 95.53 / 81.27 & 90.50 / 60.11 & 86.43 / 46.43 \\
    TCN & 95.49 / 81.09 & \textbf{90.58 / 60.32} & \textbf{86.57 / 46.75} \\
    \bottomrule 
  \end{tabular}
\end{table}

A comparison of LSTM and TCN to GBDT for three different time bins (with the default event occurring in the near term, mid term, and long term) is shown in Table 2. While GBDT seems to outperform the sequential model in predicting defaults in the near term (within 6 months after the most recent transaction), the historical information incorporated in the sequential model helps predict defaults in medium term (between 7-12 months after the most recent transaction) and long term (between 13-18 months after the most recent transaction) more effectively, indicating that our approach improves early risk detection.

\subsection{Sequence Length}
Performance of the sequential models was also dependent on the number of input transactions used (i.e., the sequence length). In Figure \ref{fig:seqlen}, we show the degradation of performance in terms of Gini and recall as the number of monthly historical transactions included in the sequences decreased, meaning the "look-back" period was shorter than a full year. Both the LSTM and TCN achieved the best results with 12 months of transactions, with TCN outperforming LSTM at each sequence length. Given the linear increase in performance as the sequence length increased, this suggests that looking back even further into borrowers' financial history might provide better performance. However, we were limited by the available data that does not predate 2016.

\subsection{Online Learning}
Since model performance is known to deteriorate with changes in consumer behavior and economic conditions, we also tested the LSTM's and TCN's ability to adapt to incoming data via online learning. To do this, we progressively tuned the weights of each model using sequences collected from three future months in 2017. We compared this method to the standard random weight initialization method in Figure \ref{fig:online}, where the weights of the model are set to small random numbers before training. 

As expected, performance gradually improved as more recent data were used to create the predictions. Additionally, progressively tuning the weights with incoming data outperformed the standard random weight initialization approach.

\subsection{Training and Inference Time}
Although the TCN outperformed the LSTM in terms of prediction performance, it is also important to consider their performance in terms of training and inference time for use in a production environment. Using an NVDIA Tesla V100 GPU, it took an average of \textasciitilde30ms to train the TCN on a single mini-batch of 512 card members, compared to \textasciitilde50ms for the LSTM. This was expected because the input sequence is processed as a whole by the TCN instead of sequentially as in the LSTM.

Anecdotally, inference time for the LSTM should be faster than the TCN. As suggested by Bai et al. \cite{Bai2018}, the LSTM only needs to save the next-to-last hidden state to memory and take in the current/incoming transactional data in order to generate a prediction. In contrast, the TCN needs to process the entire sequence. However, both methods were able to process a sequence in <1ms, which is much less than the time required for inference in our production environment (<10ms). In future work, we will analyze the full pipeline for inference that includes the time it takes to preprocess the data for the incoming transaction in addition to the time it takes load the historical data (TCN) and the next-to-last hidden state (LSTM).

\section{Conclusions}
Current models for monitoring credit risk typically utilize boosted decision
trees for their superior generalization accuracy when compared to that of other popular techniques, such as the multilayer perceptron. However, this approach is limited by its input features and inability to process sequential data efficiently.

In this paper, we investigated sequential deep learning methods for credit risk scoring and proposed a novel sampling method to generate sequences from one year of transaction-related tabular financial data. We compared the performance of our approach to the current tree-based production model and showed that our TCN, when combined with our sampling technique, achieved superior performance in terms of financial savings and early risk detection. Finally, we provided evidence that this framework would be suitable in our production environment and for online learning. 

One major concern with our approach is the lack of interpretability that plagues black-box models such as the proposed LSTM and TCN. However, since GBDT also suffers from a similar interpretability problem, we do not directly use the prediction scores of the model to make account-level decisions. Instead, established and strategic business rules are used on top of (and in parallel with) the prediction score to ensure fairness and customer satisfaction. 

Future work will focus on stress testing our deep sequential models and probing the prediction scores to determine if there exists sub-populations of card members for whom our new approach performs better than the benchmark on. It is also important to note that our data were collected during a relatively stable economic period that did not include any major recessions. As more data are collected going forward, we will be able to test the stability and generalization capability of our approach.

\bibliographystyle{ACM-Reference-Format}
\bibliography{dl_credit_risk.bbl}

\appendix

\end{document}